# Virtual local area network over HTTP for launching an insider attack

Yüksel ARSLAN[1]

**Abstract**

Computers and computer networks have become integral to virtually every aspect of modern life, with the Internet playing an indispensable role. Organizations, businesses, and individuals now store vast amounts of proprietary, confidential, and personal data digitally. As such, ensuring the security of this data from unauthorized access is critical.

Common security measures, such as firewalls, intrusion detection systems (IDS), intrusion prevention systems (IPS), and antivirus software, are constantly evolving to safeguard computer systems and networks. However, these tools primarily focus on defending against external threats, leaving systems vulnerable to insider attacks. Security solutions designed to mitigate risks originating from within the organization are relatively limited and often ineffective.

This paper demonstrates how a Local Area Network (LAN) can be covertly exposed to the Internet via an insider attack. Specifically, it illustrates how an external machine can gain access to a LAN by exploiting an unused secondary IP address of the attacked LAN, effectively bypassing existing security mechanisms by also exploiting Hyper Text Transfer Protocol (HTTP). Despite the presence of robust external protections, such as firewalls and IDS, this form of insider attack reveals significant vulnerabilities in the way internal threats are addressed.

## 1. Introduction

Information system attacks can be broadly categorized into four groups: organized, non-organized, insider, and outsider attacks [1]. Insider attackers, who may be current or former employees, business partners, or contractors, represent a particularly serious threat. These individuals often have access to network resources either in the present or from past engagements and possess detailed knowledge of internal company policies, processes, and applications. Insider attackers may also collaborate with external malicious actors to facilitate security breaches [2].

While traditional security systems such as firewalls, Intrusion Detection Systems (IDS), and Intrusion Prevention Systems (IPS) are primarily designed to protect networks from outsider threats, attackers increasingly exploit the privileges and access rights of insiders to circumvent these defenses. As insider threats become more prevalent, there is growing recognition that the most significant risks to an organization's security may originate from within. For instance, a recent study reported an increase in organizations experiencing insider attacks from 66% in 2019 to 76% in 2024 [3], [4].

Malware such as trojan horses, worms, and spyware are often employed to carry out attacks from within a network. These programs can be easily introduced to a company's computers via email attachments or through malicious websites. This paper examines the methods by which a seemingly secure LAN can be compromised by an insider attack. Specifically, it demonstrates how Virtual Private Network (VPN) technology, in conjunction with vulnerabilities in the Windows Operating System (OS) and the Transmission Control Protocol/Internet Protocol (TCP/IP) stack, can be exploited to bypass security mechanisms such as firewalls, IDS, and IPS.

To this end, we designed and implemented server and client software to execute an attack, wherein an internet-connected machine can be transformed into one connected to the target LAN. This foothold allows an attacker to perform additional attacks more easily within the compromised network. Given that Windows XP, an OS no longer supported by Microsoft, is used in this research as the LAN OS, it highlights how legacy systems remain a key vulnerability in organizational networks.

This paper is structured as follows: Section 2 gathers some related work, Section 3 provides a general overview of the LAN architecture and associated security devices. Section 4 explores the specific components of the Windows OS that are leveraged for the attack. Section 5 details the development and functioning of the client and server software used in the attack. Finally, Section 6 presents the results and discusses key findings.

## 2. Literature review

This section synthesizes existing research on insider threat detection and prevention mechanisms, and exploitation of HTTP and network broadcast traffic by malicious actors. Because our paper also exploits the same subjects.

Insider threats, due to the authorized nature of the access involved, present a unique challenge in cybersecurity. The growing sophistication of these threats has led to a proliferation of research into detection and prevention techniques, many of which are grounded in behavior analysis and anomaly detection. Al-Mhiqani et al. [5] discuss the complexity of insider threats and how behavior-based detection methods can offer promising approaches to identifying malicious activities. These methods analyze deviations from typical user behavior, which may signal unauthorized activities such as data exfiltration or sabotage. Behavior-based detection can rely on supervised and unsupervised machine learning algorithms that classify unusual actions as potential threats. Furthermore, signature-based detection approaches also continue to play a role in identifying known malicious behavior patterns by matching observed actions with predefined threat signatures [5]. Although behavior-based approaches are increasingly effective, they require large datasets to train models capable of distinguishing between normal and anomalous activities.

[1] Yüksel Arslan yuksel.arslan@ankarabilim.edu.tr
Department of Software Engineering, Ankara Science University, Ankara, Turkey

The field of big data analytics has contributed significantly to this area, providing advanced tools for processing vast amounts of network traffic data, user logs, and activity metrics. Gheyas and Abdallah [6] conducted a systematic review of these detection techniques, revealing that mostly used machine learning algorithm is Bayesian Networks followed by Support Vector Machines, Fuzzy Inference Systems, Gaussian Mixture Models, K-nearest neighbor algorithm, game theoretic approach, and regression-base models [6].

HTTP is one of the most common protocols used in network communication, and its ubiquity makes it a frequent target for attackers looking to blend malicious activities with legitimate traffic. Guofei Gu's study on botnets, for example, illustrates how command and control (C&C) channels often exploit HTTP traffic to evade detection by security systems. Botnet communication, when encapsulated within HTTP, becomes harder to detect because it mimics regular user traffic. This makes distinguishing between legitimate HTTP requests and potentially malicious traffic a significant challenge for network security tools [6]. Further exploration into the role of HTTP in malware activity was conducted by Rossow et al. through the development of Sandnet, a network traffic analysis tool that identifies the presence of malware on a network by analyzing protocol-level behaviors. Their findings revealed that DNS and HTTP were the most frequently exploited protocols, largely due to the minimal scrutiny placed on these high-volume traffic streams [7].

Another major vulnerability in modern LANs is the use of broadcast traffic, which can unintentionally expose critical information to potential attackers. Ullah et al. [9] highlight that broadcast traffic, which is often employed in various networking protocols for device discovery and service advertisements, frequently contains sensitive information that can be intercepted and exploited. While broadcast traffic facilitates network efficiency, it also poses a significant security risk when leveraged by insider threats. Attackers can exploit broadcast messages to gain access to network configurations or sensitive communications, thereby broadening their attack surface [9].

### 3. LAN description in this study

LANs are typically secured behind a firewall, as shown in Figure 1. Internet access for devices within the LAN is generally facilitated through a single public Internet Protocol (IP) address shared by all connected devices. Internally, the LAN uses private IP addresses, which are not routable over the public internet. These private addresses are reserved exclusively for communication between devices within the LAN and cannot be used to initiate direct communication with external networks.

Firewalls enforce stringent rules designed to protect the LAN from external threats. By default, they permit outbound connections from internal devices to external networks (e.g., the internet) but block unsolicited inbound connections. Outbound connections are restricted to specific Transmission Control Protocol (TCP) or User Datagram Protocol (UDP) ports, with modifications to these rules controlled by network security administrators. This makes it highly challenging to directly access or compromise a device within the LAN from outside the network.

Figure 1 illustrates the multiple security layers positioned between an external attacker and a protected LAN, emphasizing the inherent difficulties in launching an external attack on internal network resources.

The aim of this study is to demonstrate how an attacker can circumvent these protective mechanisms and gain access to all resources on a LAN protected by a firewall. While such access can typically be achieved through the use of VPN software, doing so requires specific configurations on the firewall and router, which are generally restricted to system administrators [10]. In contrast, this study presents a method by which an ordinary LAN user, without administrative privileges, can perform an insider attack using specially developed software.

Traditional VPN software typically provides a graphical user interface (GUI) to display network resources available to the user. The software developed for this study utilizes the Windows XP operating system's native GUI to display network resources to the attacker. Once deployed, the attacker gains access to all computers and shared resources within the target network via the Windows XP "Network Neighborhood" tool, simulating the appearance of a legitimate user connected to the LAN.

Importantly, the developed software requires no alterations to the firewall, router, or Intrusion Detection/Prevention Systems (IDS/IPS), allowing the attacker to bypass the need for administrative access. This ability to circumvent conventional security measures highlights the vulnerabilities posed by insider threats and legacy systems.

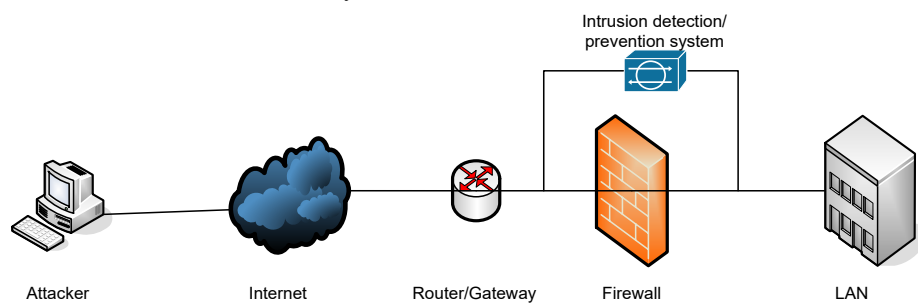

**Figure 1 Generalized Local Area Network Devices Connection**

## 4. Windows XP OS-Based LAN Functionality

The physical connection of computers within a LAN is established via a switch or hub, which operates at Layer 1 (physical layer) and Layer 2 (data link layer) of the Open Systems Interconnection (OSI) model. To facilitate resource sharing and manage network operations, the Windows XP operating system (OS) provides various utilities. In this section, we examine how Windows XP OS supports efficient LAN functionality by focusing on its TCP/IP configuration and the Computer Browser Service.

### 4.1 TCP/IP configuration in Windows XP

Windows XP OS allows multiple IP addresses to be assigned to a single network interface, enabling the system to handle multiple network communications efficiently. When a computer is configured with multiple IP addresses, the OS processes incoming and outgoing packets by checking each assigned IP address sequentially. For instance, when sending a packet such as a ping request, the OS first compares the destination IP address with the network address of the primary IP address. If there is no match, the OS proceeds to check subsequent IP addresses. If a match is found, the packet is sent to that IP address using the appropriate network interface. In cases where no match is found, the packet is forwarded to the configured gateway IP address for routing outside the local network.

This feature of assigning multiple IP addresses can be exploited by an attacker's computer (referred to here as the server or attacker computer). In such an attack, the attacker configures their computer with two IP addresses: one private IP address matching the network address of the targeted LAN and another IP address for internet connectivity, a public IP address. This dual-IP configuration enables the attacker to establish communication with both the attacked LAN and the external internet, providing a means to conduct malicious activities on the network.

### 4.2 Windows XP OS Computer Browser Servis [11]

The Windows XP "Network Neighborhood" feature provides users with a graphical interface that displays all the computers connected to the LAN. By selecting a computer, users can also view shared resources available on that machine. This functionality is enabled by the Windows XP Computer Browser Service, which operates through the broadcast of network information.

Each computer on the LAN periodically broadcasts its computer name and a list of its shared resources. A designated "main browser" aggregates this information and responds to network resource requests from other computers. The main browser also periodically announces its status, ensuring that all other computers on the network recognize its role. This service allows users to easily browse available computers and shared resources within the network.

However, the broadcast-based nature of this service introduces a vulnerability. The client software developed for this study exploits the broadcast packets sent by the Computer Browser Service. Specifically, the malicious insider's client computer intercepts these broadcast packets and transmits them to the attacker over the internet. This unauthorized access allows the attacker to gain detailed information about the network's resources and computers, facilitating further attacks.

## 5. A Virtual Local Area Network over HTTP

### 5.1 Design and Implementation

This section details the design and implementation of two software programs: the client application, which operates within the compromised network, and the server application, which runs on the attacker's machine. The client can be executed by an insider or through phishing techniques, and once activated, the server integrates into the compromised network as if it were any other connected device.

#### 5.1.1. Client Software

The client software operates on any network-connected computer and serves as an intermediary between the attacker and the LAN. It processes incoming traffic by decapsulating packets received from the attacker, stripping off the headers added by the server to bypass firewall protections. After decapsulation, the software forwards the packets to the LAN (as depicted in Figure 5). Simultaneously, it captures broadcast packets from the LAN (excluding those originating from the server itself) and packets addressed to the server. These packets are encapsulated before being transmitted to the server. The encapsulation process involves adding the appropriate TCP, IP, and Ethernet headers to the packet, effectively creating a tunnel between the LAN and the server (illustrated in Figure 2).

To facilitate packet capture and transmission, the client software leverages the open-source packet capture library WinPcap [12]. This library allows for direct interaction with the Ethernet adapter, bypassing the default Windows XP TCP/IP stack.

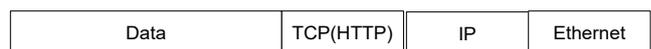

| Data | TCP(HTTP) | IP | Ethernet |

**Figure 2** Headers added to packet from client to server

The client software consists of three core components: Connection establishment, encapsulation /decapsulation and sniffer modules.

##### 5.1.1.1 Connection Establishment Module:

The client software initiates a connection to the server via TCP port 80 using the Indy TCP Client library [13]. The connection is established through the standard TCP 3-way handshake protocol (illustrated in Figure 3). During this process, the firewall records the source and destination IP addresses, along with the respective port numbers. When subsequent packets from the server are received, the firewall compares the recorded values to ensure the IP addresses and port numbers match. If they do, the packets are allowed to traverse the firewall and enter the LAN [14].

Client port number is also important. Statically configured port numbers may be changed by anti-virus/firewall software. During the connection establishment process sniffer module captures the port number assigned by the Windows XP OS and informs the encapsulation/decapsulation module. After connection establishment the work of connection establishment module finishes.

As shown in Figure 3, the connection establishment process follows these steps:

**1.** In the SYN 1 packet, the client sends a sequence number (SEQ) to the server, randomly chosen between 0 and $0 - (2^{16}-1)$.

**2.** In the SYN 2 packet, the server acknowledges the client's SEQ number by sending an acknowledgment number (ACK = SEQ + 1) and provides its own SEQ number.

**3.** Finally, the client responds with an ACK to confirm receipt of the server's SEQ number.

At this point, the TCP connection is fully established. In this implementation, the Indy TCP Client and Indy TCP Server modules are used to handle these operations, utilizing the native Windows XP TCP/IP stack [13].

During the connection process, the sniffer module monitors network traffic to capture critical connection details, such as the dynamic TCP port number assigned by the Windows XP operating system. This information is relayed to the encapsulation/decapsulation module to ensure that the correct port numbers and SEQ/ACK numbers are used in communication. Any mismatch in these values results in the firewall blocking the connection.

Once the connection is established, the role of the connection establishment module concludes, while the sniffer module continues monitoring and relaying key connection parameters.

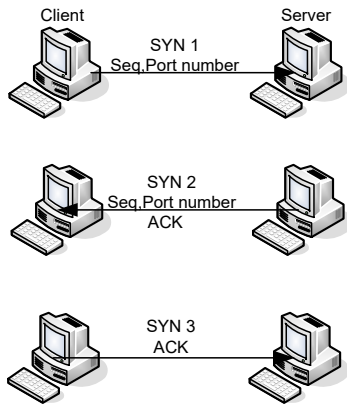

**Figure 3 Connection establishment (TCP 3-way handshake) [15]**

### 5.1.1.2 Encapsulation/Decapsulation Module:

Upon successful connection establishment, this module is activated and notified by the connection establishment module. It manages the encapsulation and decapsulation of packets received from the sniffer module, depending on their origin and destination.

The client application processes two types of packets: those originating from within the LAN and those coming from the Internet (specifically from the server).

Case 1: For packets originating from the LAN and addressed either to the server (with the server's second IP address, 192.168.0.14) or broadcast packets (with the destination Ethernet address FF:FF:FF:FF:FF), the client software adds the necessary TCP, IP, and Ethernet headers as outlined in Table 1. To avoid network loops, the client software ensures that broadcast packets originating from the server are not sent back to the server. After encapsulation, the packet is forwarded to the LAN, where it will be received by the LAN's gateway (router). The gateway checks its forwarding table to determine the appropriate interface to transmit the packet. If only one interface exists (the default interface), the gateway sends the packet to the Internet. Before reaching the gateway, however, the firewall intercepts the packet. The firewall inspects the packet to verify the established connection by referencing its network address translation (NAT) table and confirming the sequence (SEQ) numbers. Figure 4 illustrates the handling of packets from the LAN.

Case 2: The second type of packet processed by the client software are those originating from the Internet, sent by the server (attacker). These packets have the server's primary IP address (e.g., 195.212.102.201). Upon receiving such packets, the client software decapsulates the headers added by the server software and forwards the remaining data to the LAN. This data is then delivered to the intended recipient in the LAN, as specified by the server. Figure 5 demonstrates how packets from the server are handled and transmitted to the LAN after decapsulation.

An essential task of the encapsulation/ decapsulation module is handling fragmentation and defragmentation. In Ethernet networks, the maximum transmission unit (MTU) is 1500 bytes, meaning packets larger than this must be fragmented [13]. If a packet received by the client software exceeds 1460 bytes (due to added TCP and IP headers shown in Table 1), it is fragmented. The client software utilizes the TCP window field to manage fragmentation. If the window value is 1, the packet is marked as fragmented and stored until the remaining fragments arrive. Packets are divided at most into two fragments. The encapsulation/decapsulation module checks if a packet is fragmented, stores it, and reassembles it once the second fragment has been received before forwarding it to the LAN.

### 5.1.1.3 Sniffer Module:

The Sniffer module is responsible for capturing all network traffic transmitted to and from the client computer. It forwards these captured packets to the encapsulation/decapsulation module for further processing. During the connection establishment phase, the Sniffer module extracts critical parameters such as sequence (SEQ) numbers and TCP port values, which it relays to the encapsulation/decapsulation module to ensure proper packet handling and prevent network issues.

The Sniffer module leverages the WinPcap application programming interface (API) [12] to directly interface with the Ethernet adapter, bypassing higher-level network stacks and allowing for efficient packet capture at the data link layer. By capturing both incoming and outgoing traffic, the Sniffer module ensures that the client software can inspect and manipulate all relevant packets. However, this introduces the potential for packet looping. To address this, the encapsulation/decapsulation module is designed to prevent loops from occurring during packet transmission.

**Table 1 TCP, IP and ethernet header values encapsulated by the client software**

| ETHERNET | |
|---|---|
| Ethernet destination address (MAC) | MAC address of the gateway (router) in LAN. |
| Ethernet source address (MAC) | MAC address of the client computer. |
| Ethernet Protocol | 0x0800 shows ethernet protocol is the previous protocol |
| **IP** | |
| IP version | 0x45 |
| IP service | 0 |
| IP length | 40 + length of payload in this packet |
| IP ident | X random number |
| IP flags + offset | 0x4000 |
| IP time to live | 0x80 |
| IP protocol | 0x06 shows that the protocol is TCP |
| IP checksum | Routers check this value if it is not correct, they drop the packet. It is calculated by the client software. |
| Source IP address | 192.168.0.108 (client IP address) |
| Destination IP address | 195.212.102.201 (server Internet IP address) |
| **TCP** | |
| TCP source port | It is captured by the sniffer module during connection establishment |
| TCP destination port | 0x5000 |
| TCP sequence number | Next sequence no= sequence no + length of payload of previous packet |
| TCP ACK number | ACK number = sequence number of previous packet + length of payload of previous packet |
| TCP length,resv,flags | 0x5018 |
| TCP window | The client is just relaying data, so flow control done by the server OS and the computer on the LAN. We used this field for fragmentation. 0: no fragmentation, 1: fragmented |
| TCP checksum | 0x06d8 entered as a constant value, we do not check |
| TCP urgent pointer | 0 |

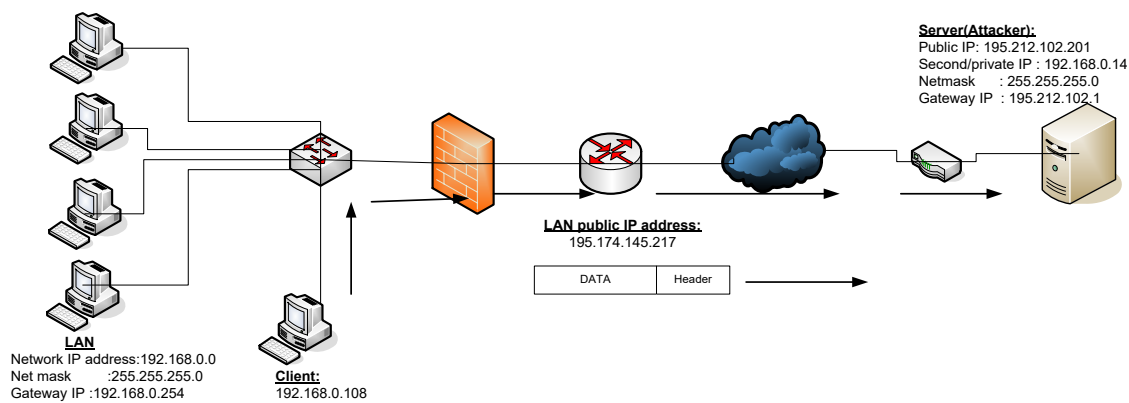

**Figure 4 A packet being received from the LAN and sent to attacker (server)**

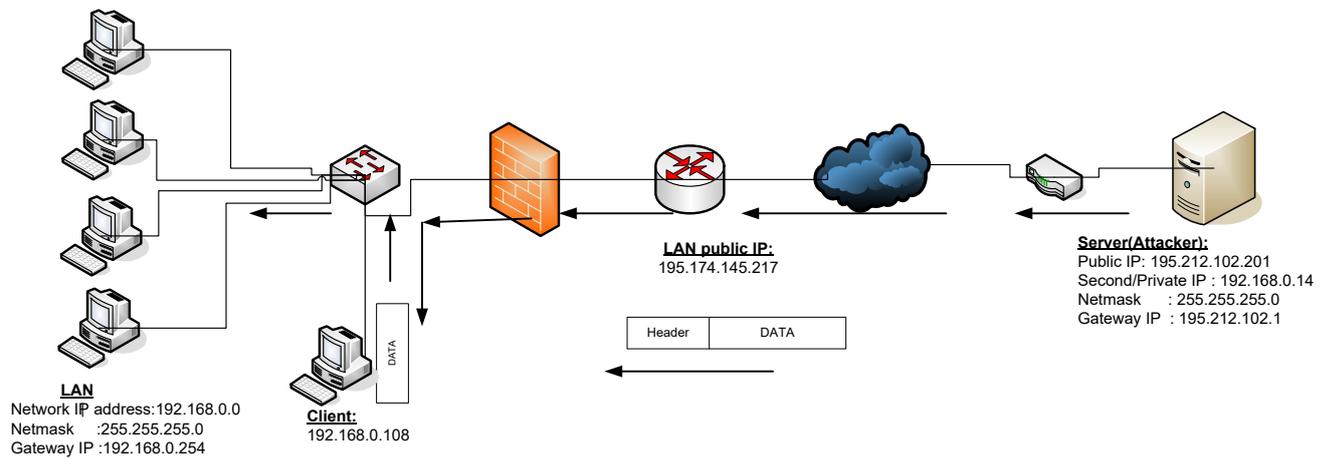

**Figure 5 A packet from the attacker being received and sent to the LAN**

### 5.1.2. Server (attacker) software

The server software runs on an Internet-accessible computer using the Windows XP operating system. For communication, the server's TCP port 80 must be open and reachable from the Internet. The server connects to the Internet via Ethernet, ensuring compatibility with the network protocols used in the attack. Once the attack is successful, the server gains visibility into all computers within the compromised LAN and their shared resources through the Windows XP network neighborhood application.

As previously discussed, Windows XP allows the configuration of multiple IP addresses for a single Ethernet interface. The server must configure a second IP address that falls within the address range of the targeted LAN. This IP address must be unique within the LAN to avoid address conflicts. To receive packets from computers in the targeted LAN, the server's IP address must align with the LAN's network prefix, such as 192.168.0.0/16. This ensures proper routing and prevents IP address collisions.

The server software is configured with key network parameters, including the IP and MAC addresses of the client and server machines, as well as the gateway address. The client's IP address is the public-facing IP of the attacked LAN, which in the case depicted in Figures 4 and 5, is 195.174.145.217.

The server software is structured similarly to the client software, consisting of three main components: HTTP server module, encapsulation/decapsulation module and sniffer module.

### 5.1.2.1 HTTP server module:

The HTTP server module is developed using Delphi 7.0 and the Indy TCP Server component [13]. It listens for incoming connection requests on TCP port 80 (HTTP). When a client initiates a connection by sending a SYN 1 packet, the server responds with a SYN 2 (SYN-ACK) packet, completing the TCP 3-way handshake. Once the connection is successfully established, the responsibility of the HTTP server module concludes, and control is handed over to the encapsulation/decapsulation module, which takes over the further processing of network traffic.

### 5.1.2.2 Encapsulation/Decapsulation module:

After the connection is established, the HTTP server module notifies the encapsulation/ decapsulation module, which then begins processing packets captured by the sniffer module. This module inspects incoming packets and performs the appropriate encapsulation or decapsulation operations.

Two primary packet types are handled: packets from the Internet (originating from the client) and packets generated by the server intended for the attacked LAN.

Case 1: If a packet originates from the client, with the client's IP address (the shared public IP address used by the entire attacked LAN), the module decapsulates the packet by removing the IP and TCP headers added by the client. The remaining payload, which is a full LAN packet, is then sent to the server's Ethernet interface. The destination IP address of this packet may correspond to the server's second IP address (assigned from the attacked LAN's private address space), or it could be a broadcast packet. The Windows XP operating system processes this packet as if it were received from a computer on the LAN. For example, in response to a ping request from the LAN, Windows XP responds as a regular network node, or in the case of a network browser packet, it displays shared resources in the network neighborhood. Figure 6 illustrates the decapsulation process, where a packet from the client is stripped of its headers and passed through the Ethernet card as if received from another LAN node. The operating system compares the destination IP address with its configured second IP (192.168.0.14) and processes the packet accordingly.

Case 2: The second type of packet is one that the server sends to the client. If the destination IP address belongs to the attacked LAN's address space or is a broadcast packet (excluding LAN-originated broadcasts to prevent loops), the module encapsulates the packet with TCP, IP, and Ethernet headers as detailed in Table 2. If any header data is missing, it is filled in using the information from Table 1.

Similar to the client software, this module is also responsible for handling fragmentation and defragmentation. When headers are added to a packet, its total size may exceed the maximum segment size, necessitating fragmentation into smaller packets. If a fragmented packet is received, the module must reassemble the fragments before forwarding the data to the server's Ethernet interface. This is achieved by storing the first fragment and combining it with subsequent fragments before processing.

### 5.1.2.3 Sniffer Module:

It works as in the client software.

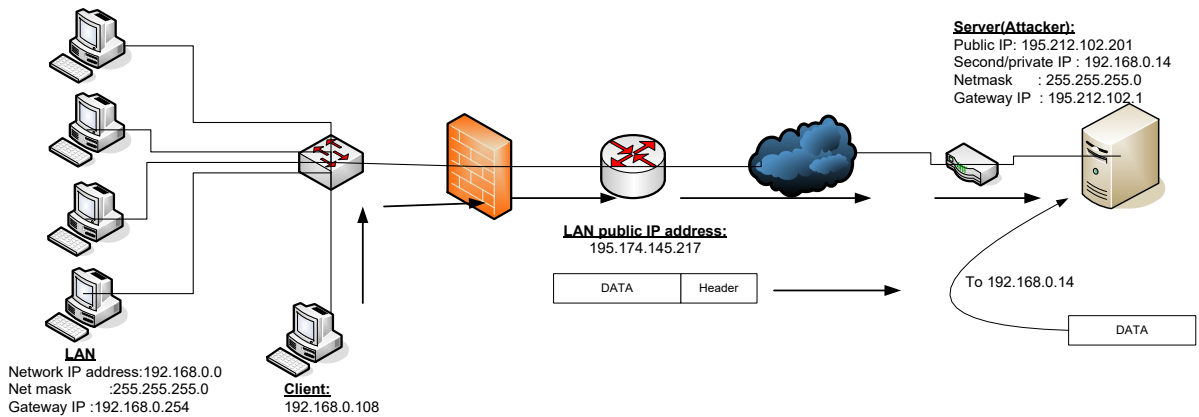

**Figure 6 A packet coming from attacked LAN being received by the attacker and sent to second IP**

**Table 2 TCP, IP and ethernet header values encapsulated by the server software. Missing values are as in Table 1**

| ETHERNET | |
|---|---|
| Ethernet destination address (MAC) | MAC address of the gateway of the server computer. |
| Ethernet source address (MAC) | MAC address of the server computer |
| **IP** | |
| Source IP address | 195.212.102.201 (Server computer Internet IP address) |
| Destination IP address | 195.174.145.217 (Attacked LAN public/Internet IP address) |
| **TCP** | |
| TCP source port number | 0x5000 |
| TCP destination port number | It is captured by the sniffer module during connection establishment. |

**5.2. Some key points explaining the inner working of client and server software.**

1- TCP Connection via Port 80: The client software initiates a connection to the server software through TCP port 80 (HTTP). This port is typically open for outgoing connections in most Local Area Networks (LANs), ensuring that client-initiated connections can pass through firewalls without being blocked. The connection is initiated from inside the LAN that is being targeted for the attack.

2- Bypassing Windows XP TCP/IP Stack Restrictions: Windows XP's TCP/IP stack does not allow TCP connections to be established outside its own stack. If Windows detects a TCP packet that is outside its stack, it automatically sends a reset packet to the destination. To circumvent this limitation, the client software establishes the initial connection using the native Windows XP TCP/IP stack. The Indy TCP Server module is used on the server side, and the Indy TCP Client module on the client side during this connection phase. The sniffer module captures critical connection parameters during this process, including sequence, acknowledgment, and port numbers. Establishing this inside-to-outside connection prevents the firewall from blocking subsequent packets from the server destined for the client.

3- Handling Broadcast and Unicast Packets: As explained in Section 4.1.1, the client software forwards all broadcast packets from the targeted LAN, as well as packets addressed to the server. Broadcast packets naturally reach the client computer, but how do packets specifically addressed to the server reach the client? When a packet is destined for the server, it contains the server's IP and MAC addresses in the destination fields. The client software takes packets from the server and forwards them to the LAN. These packets will have the server's IP and MAC addresses in their respective source fields. The server appears to be directly connected to the same port as the client. Since Ethernet switches map MAC addresses to port numbers (unless a specific security rule blocks this behavior), the switch forwards packets based on the MAC address, disregarding the IP address. The switch can also map multiple MAC addresses to a single port.

4- Handling Maximum Ethernet Packet Size: Ethernet has a maximum packet size of 1500 bytes. The client software adds an additional 20 bytes each for the TCP and IP headers, potentially creating packets larger than 1500 bytes.

When client and server software generate packets that exceed this limit, the software fragments the packets into smaller segments and transmits them as consecutive packets.

### 5.2. Ping test

To demonstrate the attack, consider issuing a ping command from a computer with IP address 192.168.0.10 connected to the LAN shown in Figure 4. The command "ping 192.168.0.14" is directed to the second IP address of the server computer, even though no device with this address exists on the LAN. The source computer (192.168.0.10) compares the network part of the destination address with its own, assuming that the destination is within the same network. To obtain the destination MAC address, the source computer sends an ARP request packet with the destination MAC address FF:FF:FF:FF:FF . As this is a broadcast packet, it is received by all computers on the LAN, including the client machine.

The encapsulation/decapsulation module in the client machine encapsulates this packet, which is then forwarded to the Internet via the gateway. The packet's original destination IP (192.168.0.14) is encapsulated with the server's Internet IP address (195.202.102.201) before transmission. Upon receiving the packet, the server identifies it as originating from the client, decapsulates the packet, and forwards it to its Ethernet interface. Since the ARP packet is addressed to 192.168.0.14, matching the server's second IP address, the server responds with an ARP reply containing its MAC address. This packet is encapsulated again by the server and sent through the Internet back to the client.

Upon receiving the response, the client decapsulates the packet and forwards it to the LAN, where the source computer (192.168.0.10) receives the ARP reply and learns the MAC address of the server. Following this, the source computer sends ICMP echo request (ping) packets to 192.168.0.14, now with the correct MAC address in the Ethernet header. These ping packets are intercepted by the client, encapsulated, and sent to the server, repeating the process outlined for ARP packets. The server responds with ICMP echo reply packets, which are routed back to the client and then forwarded to the source computer, completing the ping operation. In Figure 7 the captured packets are seen during the ping process on the server computer. Because this test has been done on a live network, not to expose complete details of the network the IP addresses different from the ones in Figures-4,5, and 6.

**Figure 7 Ping test packet capture at the server computer**

### 6. Conclusion

This study presents the development of an insider threat scenario, demonstrating how such an attack can be executed by a disgruntled employee. In this case, an external computer is able to infiltrate a local network, appearing to function as a legitimate device within the system. The research underscores that such an attack can occur even with robust security measures in place.

Despite the presence of firewalls, antivirus software, and intrusion detection and prevention systems, these defenses primarily focus on external threats, leaving internal vulnerabilities exposed. Consequently, traditional security mechanisms are largely ineffective in mitigating insider threats.

Beyond illustrating the mechanics of an insider threat, the software developed for this study also provides valuable insights into the fundamental workings of LAN. Through experimental demonstrations, we have explained key processes involved in networking, including detailed analyses of the TCP, IP, and Ethernet protocols. By manually constructing each header field, the study highlights the intricacies of network packet structures, offering an in-depth understanding of how these protocols operate in real-world scenarios.